\documentclass[11pt]{article}
\usepackage{graphicx}

\newcommand{\BABARPubYear}    {04}

\newcommand{\BABARConfNumber} {49}
\newcommand{\SLACPubNumber}{10628}

\long\def\inst#1{\par\nobreak\kern 4pt\nobreak
    {\it #1}\par\vskip 10pt plus 3pt minus 3pt}

\usepackage{graphicx}
\usepackage{dcolumn}
\usepackage{amsmath}
\usepackage{epsfig}

\RequirePackage{xspace}

\usepackage{relsize}
\def\babar{\mbox{\slshape B\kern-0.1em{\smaller A}\kern-0.1em
    B\kern-0.1em{\smaller A\kern-0.2em R}}}

\def\epem       {\ensuremath{e^+e^-}\xspace}

\def\Kbar  {\kern 0.2em\overline{\kern -0.2em K}{}\xspace}

\def\Kz    {\ensuremath{K^0}\xspace}
\def\Kzb   {\ensuremath{\Kbar^0}\xspace}
\def\KzKzb {\ensuremath{\Kz \kern -0.16em \Kzb}\xspace}
\def\Kp    {\ensuremath{K^+}\xspace}
\def\Km    {\ensuremath{K^-}\xspace}

\def\KpKm  {\ensuremath{\Kp \kern -0.16em \Km}\xspace}

\def\Dbar    {\kern 0.2em\overline{\kern -0.2em D}{}\xspace}

\def\Dz      {\ensuremath{D^0}\xspace}
\def\Dzb     {\ensuremath{\Dbar^0}\xspace}
\def\DzDzb   {\ensuremath{\Dz {\kern -0.16em \Dzb}}\xspace}
\def\Dp      {\ensuremath{D^+}\xspace}
\def\Dm      {\ensuremath{D^-}\xspace}

\def\DpDm    {\ensuremath{\Dp {\kern -0.16em \Dm}}\xspace}

\def\Bbar    {\kern 0.18em\overline{\kern -0.18em B}{}\xspace}

\def\BB      {\ensuremath{B\Bbar}\xspace} 
\def\Bz      {\ensuremath{B^0}\xspace}
\def\Bzb     {\ensuremath{\Bbar^0}\xspace}
\def\BzBzb   {\ensuremath{\Bz {\kern -0.16em \Bzb}}\xspace}
\def\Bu      {\ensuremath{B^+}\xspace}
\def\Bub     {\ensuremath{B^-}\xspace}

\def\BpBm    {\ensuremath{\Bu {\kern -0.16em \Bub}}\xspace}

\mathchardef\Upsilon="7107
\def\Y#1S{\ensuremath{\Upsilon{(#1S)}}\xspace}

\def\FourS {\Y4S}

\mathchardef\Deltares="7101
\mathchardef\Xi="7104
\mathchardef\Lambda="7103
\mathchardef\Sigma="7106
\mathchardef\Omega="710A

\def\Deltabar{\kern 0.25em\overline{\kern -0.25em \Deltares}{}\xspace}
\def\Lbar{\kern 0.2em\overline{\kern -0.2em\Lambda\kern 0.05em}\kern-0.05em{}\xspace}
\def\Sigbar{\kern 0.2em\overline{\kern -0.2em \Sigma}{}\xspace}
\def\Xibar{\kern 0.2em\overline{\kern -0.2em \Xi}{}\xspace}
\def\Obar{\kern 0.2em\overline{\kern -0.2em \Omega}{}\xspace}
\def\Nbar{\kern 0.2em\overline{\kern -0.2em N}{}\xspace}
\def\Xb{\kern 0.2em\overline{\kern -0.2em X}{}\xspace}

\def\BR         {{\ensuremath{\cal B}\xspace}}

\def\mes        {\mbox{$m_{\rm ES}$}\xspace}

\newcommand{\tev}{\ensuremath{\mathrm{\,Te\kern -0.1em V}}\xspace}
\newcommand{\gev}{\ensuremath{\mathrm{\,Ge\kern -0.1em V}}\xspace}
\newcommand{\mev}{\ensuremath{\mathrm{\,Me\kern -0.1em V}}\xspace}
\newcommand{\kev}{\ensuremath{\mathrm{\,ke\kern -0.1em V}}\xspace}
\newcommand{\ev}{\ensuremath{\mathrm{\,e\kern -0.1em V}}\xspace}
\newcommand{\gevc}{\ensuremath{{\mathrm{\,Ge\kern -0.1em V\!/}c}}\xspace}
\newcommand{\mevc}{\ensuremath{{\mathrm{\,Me\kern -0.1em V\!/}c}}\xspace}
\newcommand{\gevcc}{\ensuremath{{\mathrm{\,Ge\kern -0.1em V\!/}c^2}}\xspace}
\newcommand{\mevcc}{\ensuremath{{\mathrm{\,Me\kern -0.1em V\!/}c^2}}\xspace}

\def\invfb   {\ensuremath{\mbox{\,fb}^{-1}}\xspace}

\def\mus  {\ensuremath{\rm \,\mus}\xspace}

\def\mus        {\ensuremath{\,\mu{\rm s}}\xspace}    

\def\ra                 {\ensuremath{\rightarrow}\xspace}
\def\to                 {\ensuremath{\rightarrow}\xspace}

\newcommand{\stat}{\ensuremath{\mathrm{(stat)}}\xspace}
\newcommand{\syst}{\ensuremath{\mathrm{(syst)}}\xspace}

\def\pep2{PEP-II}

\def\gsim{{~\raise.15em\hbox{$>$}\kern-.85em
          \lower.35em\hbox{$\sim$}~}\xspace}
\def\lsim{{~\raise.15em\hbox{$<$}\kern-.85em
          \lower.35em\hbox{$\sim$}~}\xspace}

\def\CP                {\ensuremath{C\!P}\xspace}
\def\cp                {\ensuremath{C\!P}\xspace}

\xspace

\newcommand{\jprlBase}       {Phys.\ Rev.\ Lett.\xspace}
\newcommand{\jprBase}        {Phys.\ Rev.\xspace}
\newcommand{\jplBase}        {Phys.\ Lett.\xspace}

\newcommand{\nimBaseC}       {Nucl.\ Instr.\ and Methods\xspace}

\newcommand{\zpBase}         {Z.\ Phys.\xspace}

\newcommand{\nim}       [1]  {\nimBaseC~{\bf #1}}
\newcommand{\jplb}       [1]  {\jplBase\ B~{\bf #1}}

\newcommand{\jprl}      [1]  {\jprlBase\ {\bf #1}}
\newcommand{\jprd}      [1]  {\jprBase\ D~{\bf #1}}

\newcommand{\zp}        [1]  {\zpBase\ {\bf #1}}

\def\jetset74   {\mbox{\tt Jetset \hspace{-0.5em}7.\hspace{-0.2em}4}\xspace}

\def\figurebox#1#2#3{%
    \def\arg{#3}%
    \ifx\arg\empty
    {\hfill\vbox{\hsize#2\hrule\hbox to #2{\vrule\hfill\vbox to #1{\hsize#2\vfill}\vrule}\hrule}\hfill}%
    \else
    {\hfill\epsfbox{#3}\hfill}%
    \fi}

\newcommand{\btodsk}{$B^{-}\ra  D^{*0} K^{-}$}
\newcommand{\btodsp}{$B^{-}\ra  D^{*0} \pi^{-}$}

\newcommand{\btodsh}{$B^{-}\ra  D^{*0} h^{-}$}

\newcommand{\dotokp}{$D^0\ra K^-\pi^+$}
\newcommand{\dotokppp}{$D^0\ra K^-\pi^+\pi^+\pi^-$}
\newcommand{\dotokppo}{$D^0\ra K^-\pi^+\pi^0$}
\newcommand{\dotopp}{$D^0\ra \pi^-\pi^+$}
\newcommand{\dotokk}{$D^0\ra K^-K^+$}

\newcommand{\BBb}{$B\overline{B}$}
\newcommand{\pio}{$\pi^0$}
\newcommand{\deltae}{$\Delta E$}
\newcommand{\deltaemk}{$\Delta E_K$}
\newcommand{\deltaemp}{$\Delta E_{\pi}$}
\newcommand{\Do}{D^0}

\begin{document}
{\pagestyle{empty}

\begin{flushright}
\babar-CONF-\BABARPubYear/\BABARConfNumber \\
SLAC-PUB-\SLACPubNumber \\
\end{flushright}

\par\vskip 2cm

\begin{center}
\large \bf
Measurement of the Ratio \BR({\boldmath $B^{-}\ra  D^{*0} K^{-}$})/\BR({\boldmath $B^{-}\ra  D^{*0} \pi^{-}$}) and the \CP-asymmetry of
{\boldmath $B^-\ra D^{*0}_{\CP+}K^-$} decays with the \babar\ detector
\end{center}
\bigskip

\begin{center}
\large The \babar\ Collaboration\\
\mbox{ }\\
\today
\end{center}
\bigskip \bigskip

\begin{center}
\large \bf Abstract
\end{center}
We present a study of the decays \btodsp\ and \btodsk, where the
$D^{*0}$ decays into $D^0\pi^0$, with the $D^0$ reconstructed in the
\CP\ eigenstates $K^-K^+$ and $\pi^-\pi^+$ and in the (non-\CP) 
channels $K^-\pi^+$, $K^-\pi^+\pi^+\pi^-$, and $K^-\pi^+\pi^0$. We use an 
unbinned maximum 
likelihood fit to measure the signal yields. 
Using a
sample of about 123 million \FourS\ decays into $B\overline{B}$ pairs, we measure the ratios of decay rates
\begin{displaymath}
R^{*}_{\CP+}\equiv \frac{\BR(B^-\to D^{*0}_{\CP+}K^-)}{\BR(B^-\to D^{*0}_{\CP+}\pi^-)} = 0.088\pm0.021\stat^{+0.007}_{-0.005}\syst,
\end{displaymath}
and provide the first measurements of
\begin{displaymath}
R^{*}_{{\rm non}-\CP}\equiv \frac{\BR(B^-\to D^{*0}_{{\rm non}-\CP}K^-)}{\BR(B^-\to
D^{*0}_{{\rm non}-\CP}\pi^-)} = 0.0805\pm0.0040\stat^{+0.0039}_{-0.0032}\syst,
\end{displaymath}
and of the \CP asymmetry
\begin{displaymath}
A^*_{\CP+}\equiv
\frac{\BR(B^-{\ra}D^{*0}_{\CP+}K^-)-\BR(B^+{\ra}D^{*0}_{\CP+}K^+)}{\BR(B^-{\ra}D^{*0}_{\CP+}K^-)+\BR(B^+{\ra}D^{*0}_{\CP+}K^+)}=
-0.02\pm 0.24\stat\pm 0.05\syst.  
\end{displaymath}
These results are preliminary.
\vfill
\begin{center}

Submitted to the 32$^{\rm nd}$ International Conference on High-Energy Physics, ICHEP 04,\\
16 August---22 August 2004, Beijing, China

\end{center}

\vspace{0.5cm}
\begin{center}
{\em Stanford Linear Accelerator Center, Stanford University, 
Stanford, CA 94309} \\ \vspace{0.1cm}\hrule\vspace{0.1cm}
Work supported in part by Department of Energy contract DE-AC03-76SF00515.
\end{center}

\newpage
} 

\begin{center}
\small

The \babar\ Collaboration,
\bigskip

%
B.~Aubert,
R.~Barate,
D.~Boutigny,
F.~Couderc,
J.-M.~Gaillard,
A.~Hicheur,
Y.~Karyotakis,
J.~P.~Lees,
V.~Tisserand,
A.~Zghiche
\inst{Laboratoire de Physique des Particules, F-74941 Annecy-le-Vieux, France }
% ...

\end{center}\newpage

\section{INTRODUCTION}
\label{sec:Introduction}
The study of $B^-{\ra}D^{(*)0}K^{(*)-}$ decays
will play an important role in our
understanding of \cp\ violation, as they can be used to constrain the
angle $\gamma=\arg(-V_{ud}V_{ub}^*/V_{cd}V_{cb}^*)$  of the 
Cabibbo-Kobayashi-Maskawa (CKM) matrix in a theoretically clean
way by exploiting the interference between the $b\ra c\overline{u}s$ and
$b\ra u\overline{c}s$ decay amplitudes~\cite{gronau1991}.
In the Standard Model, in the absence of $D^{0}\overline{{D}^{0}}$ mixing,
$R^{*}_{\CP\pm}/R^{*}_{{\rm non}-\CP}\simeq 1+r^2\pm2r\cos\delta \cos\gamma$, where
\begin{eqnarray}\label{eq:rstar}
R^{*}_{{\rm non}-\CP/\CP\pm}\equiv \frac{\BR(B^-\ra
D^{*0}_{{\rm non}-\CP/\CP\pm}K^-)}{\BR(B^-\ra D^{*0}_{{\rm non}-\CP/\CP\pm}\pi^-)},\nonumber
\end{eqnarray}
$r$ is the
ratio of the color suppressed $B^+\ra D^{*0}K^+$ and color
allowed $B^-\ra D^{*0}K^-$ amplitudes ($r \sim
0.1-0.3$), and $\delta$ is the
\cp-conserving strong phase difference between those
amplitudes. Furthermore, defining the direct \CP asymmetry
\begin{eqnarray}\label{eq:cpa}
A^{*}_{\CP\pm}\equiv \frac{\BR(B^-{\ra}D^{*0}_{\CP\pm}K^-)-\BR(B^+{\ra}D^{*0}_{\CP\pm}K^+)}{\BR(B^-{\ra}D^{*0}_{\CP\pm}K^-)+\BR(B^+{\ra}D^{*0}_{\CP\pm}K^+)},
\end{eqnarray}
we have:
$A^*_{\CP\pm}=\pm 2r\sin\delta\sin\gamma/(1+r^2\pm
2r\cos\delta\cos\gamma)$. The unknowns $\delta$, $r$, and $\gamma$ can be
constrained from the measurements of $R^{*}_{{\rm non}-CP}$, $R^{*}_{\CP\pm}$, and
$A^{*}_{\CP\pm}$.
The Belle Collaboration has reported $R^{*}_{{\rm non}-CP} = 0.078\pm0.019\pm0.009$ using 10.1
\invfb of data~\cite{belle}.

\section{THE \babar\ DETECTOR AND DATASET}
\label{sec:babar}
We present the measurement of $R^*_{{\rm non}-CP}$, $R^*_{\CP+}$ and
$A^{*}_{\CP+}$ performed using 113 \invfb\ of data taken at the 
\FourS\ resonance by the \babar\ detector with the \pep2\ asymmetric $B$
factory. An additional 12 \invfb\
of data taken at a center-of-mass (CM) energy 40 MeV below the \FourS\ mass was
used for background studies.
The \babar\ detector is described in detail
elsewhere~\cite{detector}. 
Tracking of charged particles is provided by a five-layer silicon
vertex tracker (SVT) and a 40-layer drift chamber (DCH). Their
identification exploits ionization energy loss in
the DCH and SVT, and Cherenkov photons detected in a ring-imaging
detector (DIRC). An electromagnetic
calorimeter (EMC), comprised of 6580 thallium-doped CsI crystals,
is used to identify electrons and photons. 
These systems are mounted inside a 1.5-T solenoidal
superconducting magnet. Finally, the Instrumented Flux Return (IFR) of the 
magnet allows discrimination of muons from other particles.
We use the GEANT4 Monte Carlo (MC)~\cite{geant4} to simulate the response
of the detector, taking into account the varying
accelerator and detector conditions. 

\section{ANALYSIS METHOD}
\label{sec:Analysis}

In this analysis \btodsh\ candidates are reconstructed, where the {\em prompt} 
track $h^-$ is a kaon 
or a pion. $D^{*0}$ candidates are reconstructed from 
$D^{*0}\ra D^0\pi^0$ decays and $D^0$ mesons from 
their decays to $K^-\pi^+$, $K^-\pi^+\pi^+\pi^-$, $K^-\pi^+\pi^0$,
$\pi^-\pi^+$, and $K^-K^+$. The first
three modes are referred to as ``non-\cp\ modes'', while the
last two as ``\cp\ modes''. Reference to the charge-conjugate decays is
implied here and throughout the text, unless otherwise stated. 

Charged tracks used in the reconstruction of $D$ and $B$ meson
candidates must have a
distance of closest approach to the 
interaction point within $1.5$ cm in the transverse plane and within 
$10$ cm along the beam axis. Charged tracks from the \dotopp\ decay
must also have transverse momenta $>0.1$
\gevc and total momenta in the CM frame
$>0.25$ \gevc. 
Kaon and pion candidates from all $D^0$
decays must pass particle identification (PID) selection
criteria based on a neural network algorithm, which uses measurements of
${\rm d}E/{\rm d}x$ in the DCH and the SVT and Cherenkov photons in the DIRC.

For the prompt track to be identified as a pion or a kaon,
we require that its Cherenkov angle  ($\theta_C$) be reconstructed 
with at least five photons. To suppress misreconstructed
tracks, while maintaining high efficiency, events with prompt tracks
with $\theta_C$ $>2$ standard deviations (s.d.) away from the 
expected values for both the kaon and pion hypothesis are discarded; this
selection rejects most protons as well. The track is also discarded if it is
identified with high probability as an electron or a muon. 

Neutral pions are reconstructed by combining
pairs of photons, with energy deposits larger
than 30 MeV in the calorimeter
that are not matched to charged tracks. The $\gamma\gamma$ invariant mass is
required 
to be in the range 122--146 \mevcc. The mass resolution for all neutral
pions is typically 6--7 \mevcc. The minimum total laboratory energy
required for the $\gamma\gamma$ combinations 
is set to 200 MeV for \pio\ candidates from $D^0$ mesons. Only \pio\ candidates 
with CM momenta in the range 70--450 MeV (which we will call soft
pions, $\pi_s$) are used to reconstruct 
the $D^{*0}$. A fit is performed to constrain the $\gamma\gamma$
mass to the nominal \pio\ mass~\cite{pdg2002}.

The $D^0$ mass resolution is 11 \mevcc for 
the \dotokppo\ mode and about 7 \mevcc\ for all other modes.
A mass-constrained fit is applied to the
$D$ candidate. 
The resolution of the difference
between the masses of the $D^{*0}$ and the daughter $D^{0}$ candidates
($\Delta M$) is typically in the range 0.8--1.0 \mevcc, depending on
the $\Do$ decay mode. 
A combined cut on the measured  
$D^0$ and soft pion invariant masses and on $\Delta M$
is also applied by means of a $\chi^2$ defined as:
\begin{eqnarray}\label{eq:chi2}
\chi^2\equiv \frac{(m_{D^0}- \overline{m}_{D^0}
)^2}{\sigma^2_{m_{D^0}}}+\frac{(m_{\pi_s}- \overline{m}_{\pi_s})^2}
{\sigma^2_{m_{\pi_s}}}  
+\frac{(\Delta M- \overline{\Delta M})^2}{\sigma^2_{\Delta M}},\nonumber
\end{eqnarray}
where the mean values ($\overline{m}_{D^0}$, $\overline{m}_{\pi_s}$,
$\overline{\Delta M}$) and the resolutions ($\sigma_{m_{D^0}}$,
$\sigma_{m_{\pi_s}}$, $\sigma_{\Delta M}$) are measured 
in the data.
Correlations between the $\chi^2$ observables
are negligible. Events with $\chi^2>9$ are rejected. 

$B$ meson candidates are reconstructed by combining a $D^{*0}$ candidate with a
high momentum charged track. For the non-\CP\ modes, the charge of the
prompt track $h$ must match that of the kaon from the $D^0$ meson decay.
Two quantities are used to discriminate between signal and background: the
beam-energy-substituted mass  
$\mes \equiv \sqrt{(E_i^{*2}/2 + \mathbf{p}_i\cdot\mathbf{p}_B)^2/E_i^2-p_B^2}$
and the energy difference $\Delta E\equiv E^*_B-E_i^*/2$, 
where the subscripts $i$ and $B$ refer to the initial \epem\ system and the 
$B$ candidate respectively, and the asterisk denotes the CM frame. 

The \mes\ distribution for \btodsh\ signal can be described by
a Gaussian function centered at
the $B$ mass and {\em does not} depend on the nature of the prompt track. 
Its resolution (about $2.6$ \mevcc) is dominated by the uncertainty of
the beam energy and is slightly dependent on the $\Do$ decay mode.  
\deltae\ {\em does} depend on the mass assigned to the
tracks forming the $B$ candidate, and on the \Dz\ momentum
resolution. 
The mass hypothesis of the prompt track used to calculate \deltae\ is denoted by a
subscript $\Delta E_h$, where $h=\pi$ or $K$.
\deltaemk\ is described approximately by a Gaussian centered at zero and with
resolution 17--18 \mev, whereas
\deltaemp\ is shifted negatively by about 50 \mev.
$B$ candidates with \mes\ in the range 5.2--5.3 \gevcc\ and with 
\deltaemk\ in the range (-100 to 130) \mev\ are selected.

Multiple candidates are found in about 10--12\%
of the selected events with two and four-body \Dz\ decays and in $17$\%
of the events with \dotokppo\ decays. 
The best candidate in each event is selected based on the  
$\chi^2$ previously defined.
 
A large fraction of the background consists of \emph{continuum} (non
\BB) events; a 
powerful set of selection criteria is needed to suppress it. The selection 
is optimized to maximize the significance of the results. In the
CM frame, this background typically has 
two-jet structure, while \BB\ events are isotropic. 
We define $\theta_T$ as the angle between the thrust axes of the $B$ candidate and of the remaining charged and neutral
particles in the event, both evaluated in the
CM frame, and signed so that its component along
the  $e^-$ beam direction is positive. $|\cos\theta_T|$ is strongly peaked
near 1 for continuum events and is 
approximately uniformly distributed  for \BB\ events. For the non-\cp\ modes, $|\cos\theta_T|$ is required to be $<0.9$ for the \dotokp\ mode, and $<0.85$ for
\dotokppp and \dotokppo\ modes for which the levels of the
continuum background are higher. For the \cp\ modes, $\cos\theta_T$
is required to be in the ranges (-0.9 to 0.85) and (-0.85 to 0.8) for the
\dotokk\ and \dotopp\ modes respectively. 
Other mode-dependent selection criteria are applied: events with 
$\cos\theta_{tD}<-0.9$ ($|\cos\theta_{tD}|>0.95$) for \dotokppp\ and
\dotokppo\ (\dotopp)
modes are rejected, where $\theta_{tD}$ is the angle between the
direction of the $\Do$\ in the laboratory and 
opposite of the direction of the kaon (pion for the \dotopp\ mode) from the $\Do$\ 
in the $\Do$\ rest frame. Furthermore the momentum of the $B$ candidate in the
CM frame is required to be $>0.22$ \gevc. Finally, to reduce
combinatorial background in the \dotokppo\ final state, only those events
that fall in the enhanced regions of the Dalitz plots, according to the 
results of the Fermilab E691 experiment~\cite{dalitz2}, are selected.
The reconstruction efficiencies, based on MC simulation, are
reported in Table~\ref{tab:fitresults}.   

According to the simulation, the main contributions to the
\BB\ background for \btodsh\ events originate  
from the decays $B^-\rightarrow D^{(*)0}\rho^-$ and $B^0\rightarrow
D^{*-}h^+$. 

An unbinned maximum-likelihood (ML) fit
is used to extract yields from the data for six candidate types, signal, 
continuum background
and \BB\ background, for each choice of prompt track in the candidate
decays \btodsh.
The fit is performed independently for each \Dz\ decay
mode.

Three quantities from each selected candidate are used as input to the fit:
\deltaemk, \mes, and the $\theta_C$ of the prompt
track. The distributions of \deltaemk\ and \mes\ for the six 
candidate types are parameterized to build the probability density functions
(PDFs) which are used in the likelihood fit. 

Correlations between the \mes\ and \deltaemk\ variables for signal events
are about 5\% 
according to the simulation.
To account for these, signal MC events are used to
parameterize the signal PDFs using a method based on \emph{Kernel
Estimation}~\cite{kernel}, which allows the description of a
two-dimensional PDF. The shapes of MC and data distributions of these
variables are in excellent agreement, although
the central values are slightly shifted (more clearly 
for the \mes\ distribution) by different magnitudes for the different modes.
Hence the
\mes\ and \deltaemk\ values of the signal MC events used for the PDF for the signal 
are shifted accordingly before
fitting the data. Systematic uncertainties associated with the statistical
errors on the shifted parameters are included in the final results. 

The \mes\ PDFs for continuum background are obtained from
off-resonance data with the standard selection criteria applied. The
\mes\ distributions are parameterized with an ARGUS threshold
function~\cite{argus}: $f(m_{ES})\sim m_{ES} \sqrt{1-y^2}\exp{[-\xi(1-y^2)]}$, 
where $y=m_{ES}/m_0$ and $m_0$ is the mean energy of the beams
in the CM frame.
The \deltaemk\ PDFs for background candidates from the continuum are well
parameterized with exponential functions whose parameters are 
determined by
fitting the \deltaemk\  distributions of the selected \btodsh\ sample
in the off-resonance data. Both the \mes\ and the \deltaemk\ PDFs for the continuum
background are taken to be the same for \btodsp\ and \btodsk\ decays. MC and
data distributions of \mes\ and \deltaemk\ obtained with looser selection
criteria, in order to increase the statistics, agree well for \btodsp\ and \btodsk\
decays, validating this 
assumption. 
The PDFs used for the \cp\ modes are the same as for the \dotokp\ mode, as
very few events from off-resonance data pass the selection criteria for these
modes. This assumption is validated by comparing the distributions of \mes\ and
\deltaemk\ for the \dotokp\ and \dotokk\ or \dotopp\ modes obtained using a
looser event selection.

The correlation between \mes\ and \deltaemk\ for the \BB\ background
is taken into account with a two-dimensional PDF determined
from simulated events, in a similar way to that used for the signal.

The PID PDFs for the
kaon and pion hypotheses of the prompt track are determined from
distributions, in bins of momentum and polar angle, of 
the difference between the reconstructed and expected 
$\theta_C$ of kaons and pions from \Dz\ decays in a control
sample which exploits the decay chain $D^{*+}\ra  
D^0\pi^+$, $D^0\ra K^-\pi^+$ to kinematically identify the tracks.

Initial PDFs are parameterized for each candidate type as described above. 
These do not describe exactly all the distributions of the observables
used in the fit and all their correlations. This is in part caused by residual
particle misidentification of the prompt tracks due to long and small 
non-gaussian tails in the $\theta_C$ residual distributions not removed by the 
prompt track selection. Hence, we use corrected PDFs for each candidate type which 
are weighted sums of all the initial PDFs. The weights are determined by 
fitting pure samples of simulated signal events and of background from
off-resonance real and MC data. The corrections affecting the signal yields are 
typically of order $1$\%.
The fractional systematic uncertainties for the signal yields associated
to these corrections are in the range 0.1--6.0\% depending on
the $\Do$ decay mode.

The likelihood $\mathcal{L}$ for the selected sample is given by the
product of the final PDFs for each individual candidate and a Poisson
factor:
\begin{eqnarray}\label{eq:pdf}
\mathcal{L}\equiv \frac{e^{-N'}(N')^N}{N!}\prod_{i=1}^{N}\sum_{j=1}^6 \frac{N_{j}}{N'}\mathcal{P}^i_j(m_{ES},\Delta E_K,\theta_C) \nonumber
\end{eqnarray}
where $N$ is the total number of events, $N_{j}$ are the yields for each
of the previously defined six candidate types, and $N'\equiv \sum_{j=1}^6 N_{j}$,
$\mathcal{P}^i_j(m_{ES},\Delta E_K,\theta_C)$ is the 
probability to measure the particular set of 
physical quantities (\mes,\deltaemk,$\theta_C$) in the $i^{th}$ event for a
candidate of type $j$. The Poisson factor is the probability of observing
$N$ total events when $N'$ are expected.
The quantity $\mathcal{L}$ is maximized with respect to the
six yields using
MINUIT~\cite{minuit}. The fit has also been performed on luminosity weighted MC 
and high statistics toy MC events and it has been found to be unbiased. The 
yields thus found are corrected to account for
small differences in resolutions for \deltaemk\ and \mes\ between data and
simulation in the parameterization of the signal.

The preliminary results of the fit are reported in detail in
Table~\ref{tab:fitresults}.  
These yields are used to determine the CP asymmetry parameters.
We measure:
\begin{eqnarray}
R^*_{\CP+} &=& 0.088 \pm0.021 \stat^{+0.007}_{-0.005} \syst, \nonumber\\
R^*_{{\rm non}-CP}       &=& 0.0805\pm0.0040\stat^{+0.0039}_{-0.0032}\syst, \nonumber\\
R^*_{\CP+}/R^*_{{\rm non}-CP} &=& 1.09\pm0.26\stat^{+0.10}_{-0.08}    \syst, \nonumber\\
A^*_{\CP+}    &=& -0.02\pm0.24\stat\pm0.05    \syst. \nonumber
\end{eqnarray}

\begin{figure}[!htb]
\begin{center}
\includegraphics[width=15.0cm]{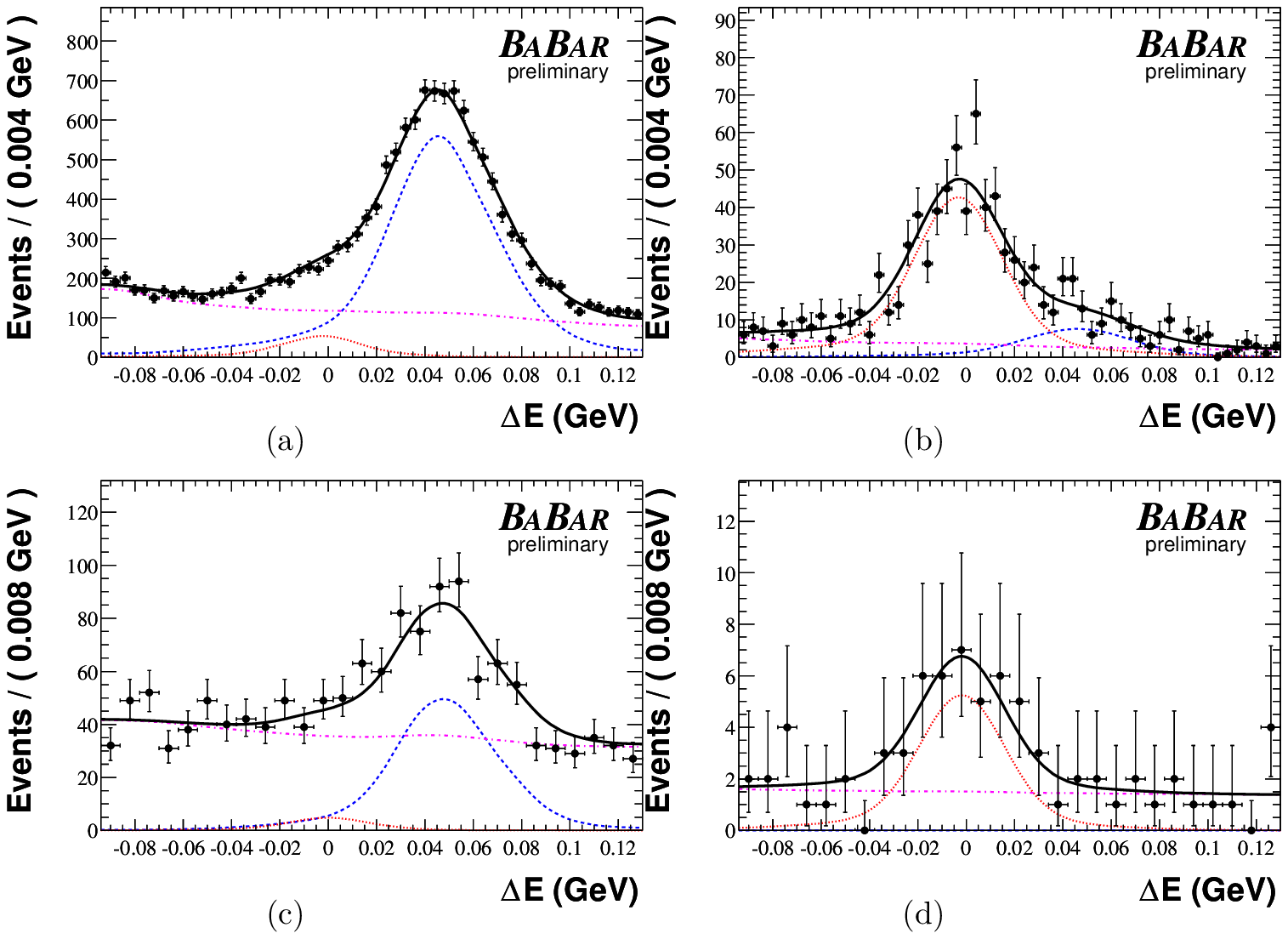}
\caption{Distributions of \deltaemk\ in the $B\rightarrow D^{*0}h$ sample, for
$D^0\rightarrow K^-\pi^+, K^-\pi^+\pi^0, K^-\pi^+\pi^+\pi^-$ (top, (a),
(b)) and  $D^0\rightarrow K^-K^+, \pi^-\pi^+$ (bottom, (c), (d)), before
(left, (a), (c)) and after (right, (b), (d)) enhancing the $B\rightarrow
D^{*0}K$ component by requiring that the prompt track be consistent with the
kaon hypothesis and $\mes>5.27$\gevcc.  The
\btodsp\ signal contribution on the right of each plot is shown as a
dashed line,  
the \btodsk\ signal on the left as a dotted line, and the background
as a dashed-dotted line. The total fit with all the contributions is
shown with a thick solid line.}
\label{fig:fit}
\end{center}
\end{figure}

Figure~\ref{fig:fit} shows the distributions of \deltaemk\ for the 
combined non-\CP\ and \CP\ modes before and after enhancing the $B\rightarrow
D^{*0}K$ component. This is accomplished by requiring that the prompt track be consistent with the
kaon hypothesis and that $\mes>5.27$\gevcc. The \deltaemk\
projections of the fit results are also shown.
 
\begin{table}[ht]
\caption{Results of the yields from the ML fit. For the
\cp\ modes the results of the fit separately for the $B^+$ and $B^-$
samples are also quoted. Errors are statistical only. The efficiencies
($\epsilon$) according to MC simulation are
also reported.}
\label{tab:fitresults}
\begin{center}
\begin{tabular}{|l|c|c|c|c|}
\hline
$D^0$ mode &\  $N(B\rightarrow D^{*0}\pi)$ &\ $N(B\rightarrow D^{*0}K)$ &\
$\varepsilon$ (\%)\\
\hline
\hline
$K^-\pi^+$           &\  $2502\pm 55$ &\  $218  \pm 17$  &\ 17\\
$K^-\pi^+\pi^+\pi^-$ &\  $3105\pm 66$ &\  $231  \pm 20$  &\  6\\
$K^-\pi^+\pi^0$      &\  $2984\pm 63$ &\  $230  \pm 21$  &\ 10\\ 
\hline 
\hline
$K^-K^+$             &\  $245 \pm 18$ &\  $21.4 \pm 5.2$ &\ 15\\
$K^-K^+$ [$B^+$]     &\  $117 \pm 13$ &\  $11.6 \pm 3.9$ &\ 15\\
$K^-K^+$ [$B^-$]     &\  $126 \pm 13$ &\  $ 9.8 \pm 3.7$ &\ 15\\
\hline
$\pi^-\pi^+$         &\  $115 \pm 14$ &\  $ 7.4 \pm 4.6$ &\ 13\\
$\pi^-\pi^+$ [$B^+$] &\  $67  \pm 11$ &\  $ 1.8 \pm 3.0$ &\ 13\\
$\pi^-\pi^+$ [$B^-$] &\  $46  \pm  9$ &\  $ 5.4 \pm 3.5$ &\ 13\\
\hline
\end{tabular}
\end{center}
\end{table}

The ratio of the decay rates for \btodsp\ and \btodsk\ is
separately calculated for the different $D^0$ decay channels and is
computed with the signal yields estimated with the
ML fit and listed in 
Table~\ref{tab:fitresults}. The resulting ratios are scaled by a
correction factor of the order of a few percent, which is  estimated with
simulated data, 
and takes into account small differences in the efficiency between
\btodsk\ and \btodsp\ event selections. The results are listed in
Table~\ref{tab:final_ratio}. 

\begin{table}[ht]
\caption{Measured ratios
for different \Dz\ decay modes. The first error is statistical, the second is
systematic.} 
\label{tab:final_ratio}
\begin{center}
\begin{tabular}{|l|c|}
\hline
\btodsh\ Mode &\ \textbf{\BR($B{\ra}D^{*0}K$)/\BR($B{\ra}D^{*0}\pi$)} $(\%)$\\
\hline
\hline
\dotokp&\ \ \ \ \ \        $9.10\pm 0.74^{+0.41}_{-0.32}$\\
\dotokppp&\ \ \ \ \ \      $7.44\pm 0.65^{+0.32}_{-0.29}$\\
\dotokppo&\ \ \ \ \ \      $7.73\pm 0.71^{+0.62}_{-0.60}$\\
\hline
Weighted Mean (non-\CP)&\ \ \ \ \ \  $8.05\pm 0.40^{+0.39}_{-0.32}$\\
\hline
\hline
\dotokk &\ \ \ \ \ \       $9.0\pm 2.3^{+0.5}_{-0.4}$\\
\dotopp &\ \ \ \ \ \       $7.5\pm 4.8^{+1.4}_{-1.2}$\\
\hline
Weighted Mean (\CP)&\ \ \ \ \ \  $8.8\pm 2.1^{+0.7}_{-0.5}$\\
\hline
\end{tabular}
\end{center}
\end{table}

\section{SYSTEMATIC STUDIES}

The sources of systematic uncertainties for the yields have been identified and their
contributions (for the measurement of $R^*_{({\rm non}-)CP}$) are reported in
Table~\ref{tab:syst_ratio}. 
Uncertainties of the signal parameterizations of \deltaemk\ and \mes\ arise
from the assumed shapes of the
PDFs and discrepancies between real and simulated data;  
all the parameters of the \deltaemk\ and \mes\ PDFs have also been varied
according to their statistical uncertainties (one s.d.) and the  
variations in the yields are taken (with their signs) as systematic
uncertainties. 
For the \BBb\ and continuum background, the
systematic uncertainties due to limited statistics of the MC and off-resonance data 
have been calculated varying the \deltaemk\ and \mes\ PDF by their
statistical uncertainties.
There are several contributions to the PID systematic uncertainty for the 
prompt track: the uncertainty
due to limited statistics is calculated by varying each
parameter of the PDF in each bin in momentum and polar angle by its
uncertainty (keeping constant all other parameters in the same bin 
and all parameters in all the other bins) and summing all the contributions 
in quadrature; results obtained with alternative PID PDFs, which account
for different 
$\theta_C$ residual
shapes and for discrepancies between data and simulation, are also included as
systematic 
uncertainties. The systematic uncertainties due to the PDF reweighting procedure 
have been evaluated. Finally, errors associated to the efficiency correction factor
are also included. 

Many of the systematic uncertainties for the signal yields have similar
effects  
on the \btodsk\ and \btodsp\ events (they increase or decrease both fractions
simultaneously), hence their effect is reduced in deriving the
systematic uncertainty for the measurement of the ratios, when all
correlations are taken into 
account. 
Overall, the main sources of systematic
uncertainties for the
measurement of both $R^*_{({\rm non}-)CP}$ and $A^*_{\CP+}$ are due to the characterization
of the shapes of \mes\ and \deltaemk for the signal, to the
characterization of the \mes\ PDFs for the background, to the particle
identification, and to the uncertainty of the PDFs weighting procedure and of the efficiency
correction factors. The systematic uncertainty for $A^*_{\CP+}$ due to
possible detector charge asymmetries is evaluated by measuring asymmetries
analogous 
to those defined in Eq.~\ref{eq:cpa}, but for \btodsp\ and \btodsk\ events (the latter
uniquely for the non-\CP\ modes), where \CP\ violation is
expected to be negligible. Results for all modes are then combined, taking
correlations into account. The measured asymmetry is
$-0.010\pm0.012\stat^{+0.002}_{-0.001}\syst$ 
and its maximum variation from zero up to one s.d. (0.022) is taken,
conservatively, as a further symmetric systematic error on $A^*_{\CP+}$.
When combining the  
results for the different modes, all systematic
and statistical uncertainties are considered to be uncorrelated, except
for the contributions of the
PID PDF (common to all modes) and of the detector charge asymmetry in the
measurement of $A^*_{\CP+}$, which are 
considered to be completely correlated. For the measurement of $R^*_{CP+}/R^*_{{\rm non}-\CP}$
all systematic uncertainties have been considered to be uncorrelated; this
assumption is conservative, and has negligible effect on the
largely statistically limited final result.  

\begin{table}[ht]
\caption{Average systematic uncertainties for $R^*_{({\rm non})-CP}$.} 
\begin{center}
\begin{tabular}{|l|c|c|}
\hline
Systematic & $R^*_{{\rm non}-CP}$ & $R^*_{CP}$  \\
Source & $\Delta R^*_{{\rm non}-CP}/R^*_{{\rm non}-CP} (\%)$& $\Delta R^*_{CP}/R^*_{CP} (\%)$\\
   &  non-\CP\ modes &  \CP\ modes   \\
\hline
\hline
$\Delta E_K $ (signal)       &$^{+  1.9}_{-  1.8}$   &$^{+  2.4}_{-  2.3}$  \\
$\Delta E_K(q\bar q)$        &$^{+  0.3}_{-  0.4}$   &$^{+  1.5}_{-  2.2}$  \\
$\Delta E_K(B\bar B)$        &$^{+  0.2}_{-  0.4}$   &$^{+  1.1}_{-  1.7}$  \\
\mes (signal)                &$^{+  0.4}_{-  0.3}$   &$^{+  0.6}_{-  0.7}$  \\
\mes $ (q\bar q)$            &$^{+  0.9}_{-  0.9}$   &$^{+  4.9}_{-  2.1}$  \\
\mes $ (B\bar B)$            &$^{+  1.5}_{-  1.6}$   &$^{+  3.4}_{-  3.3}$  \\
PDF Weights			     &$^{+  2.7}_{-  2.7}$   &$^{+  1.0}_{-  1.2}$  \\
PID PDF                      &$^{+  2.7}_{-  1.8}$   &$^{+  2.2}_{-  2.0}$  \\
$\varepsilon$ Correction     &$^{+  1.5}_{-  1.5}$   &$^{+  2.0}_{-  2.0}$  \\
\hline
\end{tabular}
\end{center}
\label{tab:syst_ratio}
\end{table}
\label{sec:Systematics}

\section{SUMMARY}
\label{sec:Summary}
In conclusion, we have measured the ratio of the decay rates for
$\BR(B^-\ra D^{*0}K^-)$ 
and $\BR(B^-\ra D^{*0}\pi^-)$, with
non-\CP\ eigenstates. This constitutes the most
precise measurement of this kind. We have also performed the first measurement of
the same ratio and of the \cp\
asymmetry $A^*_{\CP+}$ for $D^0$ mesons decaying to \cp\ eigenstates. These
results, together with measurements exploiting $B^-\ra D^{0}K^-$, $B^-\ra
D^{0}K^{*-}$ and $B^-\ra
D^{*0}K^{*-}$ decays~\cite{belle,expres}, constitute a first step
towards measuring the 
angle $\gamma$. All the results presented in this document are preliminary. 
Assuming factorization and 
flavor-SU(3) symmetry, theoretical calculations (in the tree-level
approximation) predict: $\BR(B^-\ra D^{*0}K^-)/\BR(B^-\ra
D^{*0}\pi^-)\sim(V_{us}/V_{ud})^2(f_K/f_\pi)^2\sim$0.074, where $f_K$
and$f_\pi$ are the meson decays constants~\cite{RinMS}. Our results accord
with these predictions.

\section{ACKNOWLEDGMENTS}
\label{sec:Acknowledgments}
We are grateful for the 
extraordinary contributions of our \pep2\ colleagues in
achieving the excellent luminosity and machine conditions
that have made this work possible.
The success of this project also relies critically on the 
expertise and dedication of the computing organizations that 
support \babar.
The collaborating institutions wish to thank 
SLAC for its support and the kind hospitality extended to them. 
This work is supported by the
US Department of Energy
and National Science Foundation, the
Natural Sciences and Engineering Research Council (Canada),
Institute of High Energy Physics (China), the
Commissariat \`a l'Energie Atomique and
Institut National de Physique Nucl\'eaire et de Physique des Particules
(France), the
Bundesministerium f\"ur Bildung und Forschung and
Deutsche Forschungsgemeinschaft
(Germany), the
Istituto Nazionale di Fisica Nucleare (Italy),
the Foundation for Fundamental Research on Matter (The Netherlands),
the Research Council of Norway, the
Ministry of Science and Technology of the Russian Federation, and the
Particle Physics and Astronomy Research Council (United Kingdom). 
Individuals have received support from 
CONACyT (Mexico),
the A. P. Sloan Foundation, 
the Research Corporation,
and the Alexander von Humboldt Foundation.

%

%
%


\begin{thebibliography}{99}

\bibitem{gronau1991} M.~Gronau and D.~Wyler, \jplb{265}, 172 (1991);
  M.~Gronau and D.~London, \jplb{253}, 483 (1991); D.~Atwood, I.~Dunietz
  and A.~Soni, \jprl{78}, 3257 (1997); A.~Soffer, \jprd{60}, 054032 (1999);
  M.~Gronau, \jprd{58}, 073301 (1998); Z.~Xing, \jprd{58}, 093005 (1998);
  J.H.~Jang and P.~Ko, \jprd{58}, 111302 (1998); M.~Gronau and
  J.L.~Rosner, \jplb{439}, 171 (1998).
\bibitem{belle} Belle Collaboration, K.~Abe $et\ al.$, \jprl{87}, 111801 (2001).
\bibitem{detector} \babar\ Collaboration, B.\ Aubert {\em et al.}, \nim{A479}, 1 (2002).
\bibitem{geant4} GEANT4 Collaboration, S. Agostinelli {\em et al.}, \nim{A506}, 250 (2003). 
\bibitem{pdg2002} Particle Data Group, K. Hagiwara $et\ al.$, \jprd{66}, 010001 (2002)
\bibitem{dalitz2} E691 Collaboration, J.~C.~Anjos {\em et al.}, \jprd{48}, 56 (1993)
\bibitem{kernel} K.~S.~Cranmer, Comp. Phys. Commun. {\bf 136}, 198 (2001).
\bibitem{argus} ARGUS Collaboration, H. Albrecht $et\ al.$, \zp{C48}, 543 (1990)
\bibitem{minuit} F. James, Comput. Phys. Commun. {\bf 10}, 343 (1975).
\bibitem{expres} \babar\ Collaboration, B.\ Aubert {\em et al.}, \jprd{69}, 051101 (2004);
\babar\ Collaboration, B.\ Aubert {\em et al.}, \jprl{92}, 202002 (2004); \babar\ Collaboration, B.\ Aubert {\em et al.}, \jprl{92} 141801 (2004);
Belle Collaboration, S.K.~Swain {\em et al.}, \jprd{68}, 051101 (2003).
\bibitem{RinMS} M~Gronau {\em et al.}, \jprd{52}, 6356 (1995).

\end{thebibliography}
\end{document}